\pdfoutput=1
\documentclass[final]{elsart3p}
\usepackage{graphicx}
\usepackage[center]{subfigure}

\newcommand{\noun}[1]{\textsc{#1}}
\newcommand{\deriv}[2]{\ensuremath{\frac{\partial #1}{\partial #2}}}

\newcommand{\hthe}{\ensuremath{h_\theta}}
\newcommand{\Bp}{\ensuremath{B_\theta}}
\newcommand{\Bt}{\ensuremath{B_\zeta}}
\newcommand{\Vec}[1]{\ensuremath{\mathbf{#1}}}
\newcommand{\bvec}{\Vec{b}}

\newcommand{\Bvec}{\Vec{B}}

\newcommand{\Jvec}{\Vec{J}}

\newcommand{\sbt}{\ensuremath{\sigma_{B\theta}}}

\newcommand{\Mat}[1]{\ensuremath{\mathcal{#1}}}

\newcommand{\link}[1]{\texttt{#1}}

\journal{Journal of Plasma Physics}

\begin{document}
\begin{frontmatter}

\title{BOUT++: Recent and current developments}
\author{B.D.Dudson\corauthref{cor}$^a$, }
\corauth[cor]{Corresponding author.}
\ead{benjamin.dudson@york.ac.uk}
\author{A.Allen$^a$, }
\author{G.Breyiannis$^b$, }
\author{E.Brugger$^c$, }
\author{J.Buchanan$^d$, }
\author{L.Easy$^{a,d}$, }
\author{S.Farley$^e$, }
\author{I.Joseph$^c$, }
\author{M. Kim$^f$, }
\author{A.D.McGann$^a$, }
\author{J.T.Omotani$^d$, }
\author{M.V.Umansky$^c$, }
\author{N.R.Walkden$^{a,d}$, }
\author{T.Xia$^{c,g}$, }
\author{X.Q.Xu$^c$}
\address{$^a$ York Plasma Institute, Department of Physics, University of York, Heslington, York, YO10 5DD, UK}
\address{$^b$ Japan Atomic Energy Agency, Rokkasho Fusion Institute, Rokkasho-mura, 039-3212, Japan}
\address{$^c$ Lawrence Livermore National Laboratory, Livermore, CA 94551, USA}
\address{$^d$ CCFE, Culham Science Centre, Abingdon, OX14 3DB, UK}
\address{$^e$ Mathematics Department, Illinois Institute of Technology, Chicago, IL 60616, USA}
\address{$^f$ Department of Physics, POSTECH, Pohang, Gyeongbuk 790-784, Korea}
\address{$^g$ Institute of Plasma Physics, Chinese Academy of Sciences, Hefei, China}

\begin{abstract}
BOUT++ is a 3D nonlinear finite-difference plasma simulation code, capable
of solving quite general systems of PDEs, but targeted particularly
on studies of the edge region of tokamak plasmas. 
BOUT++ is publicly available, and has been adopted by a growing number of researchers
worldwide. Here we present improvements which have been made to the code since its
original release, both in terms of structure and its capabilities. Some
recent applications of these methods are reviewed, and
areas of active development are discussed. We also present
algorithms and tools which have been developed to enable creation of inputs
from analytic expressions and experimental data, 
and for processing and visualisation of output results. This includes a
new tool \noun{Hypnotoad} for the creation of meshes from experimental equilibria.

Algorithms have been implemented in BOUT++ to solve a range of linear algebraic problems 
encountered in the simulation of reduced MHD and gyro-fluid models: A preconditioning
scheme is presented which enables the plasma potential to be calculated efficiently
using iterative methods supplied by the PETSc library, without invoking the Boussinesq approximation. Scaling studies are also performed of a linear solver used as part of
physics-based preconditioning to accelerate the convergence of implicit time-integration schemes.

\end{abstract}


\end{frontmatter}

\section{Introduction}

The edge region of tokamak plasmas is of crucial importance to their feasibility and economic
viability as fusion reactors. It is the interface between the hot ($\sim 1-10$ keV), fully ionised
core required for fusion, and the plasma facing material surfaces which must remain cool ($< 1$eV)
to avoid excessive damage or sputtering of impurities, which could contaminate the core plasma. 
The wide range of temperatures and hence collisionality regimes; nonlinear plasma dynamics; atomic ionisation, charge-exchange and recombination
processes; impurities; and interaction with material surfaces, make modelling the plasma edge
challenging. This is further complicated by the magnetic geometry, which is usually arranged into
an 'X-point' configuration, in which the closed magnetic surfaces of the core are surrounded by open
magnetic field-lines, which transport energy and particles leaving the core away into divertor regions designed
to handle high heat fluxes. 

Simulations of the edge of tokamaks which include many of the important atomic and impurity radiation processes
are routinely performed using 1-D~\cite{goswami2001,fundamenski2001,nakamura2011} and
2-D~\cite{rognlien1992,schneider2006,kawashima2006} codes. These however cannot correctly
predict the plasma transport across the magnetic field, which is usually anomalous (turbulent)~\cite{dippolito2011}, and for which diffusion (Fick's law) is a poor approximation~\cite{garcia-2007,naulin-2007,ghendrih2012}. The need for 
a first-principles understanding of cross-field transport in the edge region
has received increasing attention in recent years, and several new 3-D codes have been developed
to study edge turbulence \cite{ricci2012,ricci2013,tamain2010}. As discussed above, the structure 
and  dynamics of the edge region of tokamaks involves a complicated interaction between many
physical processes, and as a result it is not clear {\it a priori} which model or combination
of models is most appropriate. To minimise duplication of effort, there is a need for a flexible 
code which can be adapted to solve a range of different models, and is modular enough that it
can be extended in multiple ways by a large group of users. BOUT++~\cite{Dudson2009} is an
open-source 3D nonlinear finite difference code which aims to fill that need.

BOUT++ was developed originally to study tokamak edge plasma physics~\cite{Dudson2009}, 
taking ideas and lessons learned from the earlier BOUT code 
\cite{xu-1999,xu-2000,umansky-2006,xu-2008}. It is highly modular, operates in general curvilinear
coordinates and complicated mesh topologies, and can be applied to the solution of quite general PDEs
in three dimensions + time. 
Recent applications of BOUT++ include the study of plasma transients (Edge Localised Modes, ELMs)~\cite{xu2010,xia2012,xi2013}, plasma turbulence~\cite{friedman2012}, and the dynamics of isolated `blobs' in 3D~\cite{angus2012prl,angus2012cpp,walkden2013}.

The BOUT++ distribution is publicly available on Github\footnote{BOUT++ public distribution http://github.com/boutproject}, and comes with
a test suite and variety of plasma physics models and examples. Some have been used to produce
results published elsewhere (e.g. ELM, LAPD turbulence, and blob models), whilst others can be used as a starting point for new physics studies.
This paper describes the 2.0 release of BOUT++, which was used as a basis for the 2013 BOUT++ workshop\footnote{BOUT++ website: \link{http://bout2013.llnl.gov}}: Section~\ref{sec:structure} briefly describes modifications to the structure of BOUT++
which have been made to accommodate further development; Section~\ref{sec:capabilities} describes improvements and
new capabilities which have been added to
BOUT++ since its original release~\cite{Dudson2009}. Section~\ref{sec:tools} details the 
development of pre- and post-processing tools for equilibrium input and visualisation. 
In section~\ref{sec:conclusions} we conclude and discuss future directions for development.

\section{Code structure}
\label{sec:structure}

The BOUT++ development community has expanded significantly following its release and 2011 workshop,
and with that has come the need
to adopt more professional software development practices: Git\footnote{Git version control: \link{http://git-scm.com/}} is used for version control,
along with a system of feature branches adopted from the PETSc development group~\cite{efficient} which is described in detail on the BOUT++ development page\footnote{BOUT++ development model \link{http://http://boutproject.github.io/devel.html}}. An important addition has been a test suite which can be run quickly before
changes are committed, to check that nothing obvious has been broken. This has greatly simplified the
process of checking code, and has resulted in many bugs being caught before they could
affect production code. The majority of these tests are not physics simulations, as these would take
too long to perform and so discourage regular use. Instead, tests are designed to check individual components
independently for a range of inputs and processor configurations, so that the cause of a test failure can be quickly identified.

A more rigorous set of tests using the Method of Manufactured Solutions~\cite{roache1998,salari2000} for code verification are currently under development, and will be published
elsewhere.

\subsection{Factory pattern}
\label{sec:factory}

To enable BOUT++ to be extended, and new implementations of
solvers for boundary and initial-value problems to be added independently, components have been refactored and
organised along the Factory pattern~\cite{gamma1995,knepley2012}, a widely used method
to separate interface from implementation. Each component of BOUT++, such as file I/O or 
time integration solver, has a well defined interface (in C++ a base class with virtual members). 
Several implementations of this interface can coexist, and the user code doesn't depend 
on which implementation is used. To create a particular instance, the static member function
``create'' is called. For example, time-integration schemes implement the ``Solver'' interface,
so creating a new solver is done by the following:
\begin{verbatim}

 Solver *s = Solver::create();

\end{verbatim}
Which particular instance of Solver is created (RK4, CVODE, PETSc etc.) is set by options stored in a tree 
structure, which can be set in the input file or on the command-line. By default the options section
for the Solver class is called ``solver'', so to choose the rk4 method on the command-line
the user adds
\begin{verbatim}

solver:type=rk4

\end{verbatim}
To allow multiple solvers with different
settings to be used simultaneously, the option section can be passed during creation:
\begin{verbatim}

 Solver *s = Solver::create(
         Options::getRoot()->getSection("mysolver")
                            );

\end{verbatim}
The options for this solver will now be in the ``mysolver'' section of the options tree, and can be changed independently.

This pattern enables users to experiment with different numerical methods with minimal changes
to the program inputs. As new capabilities are added to BOUT++, such as new PDE and ODE solvers, existing
models can take advantage of them without needing to modify any code, only the input settings.
To the extent possible, this separation of interface and implementation reduces the number of dependencies
between parts of the code and allows researchers to benefit from each others work on separate components.

\subsection{Library interface}

The interface between the core BOUT++ code and problem-specific ``user'' code has also been modified
since the original publication. The original BOUT++ was structured as a framework so that the
\texttt{main()} function was defined internally, and the user supplied two functions: one for
initialisation of the desired physics model, and one which calculated the time-derivative of each 
evolving variable given the system state. Further callback functions were later added for optional preconditioning (section~\ref{sec:timeintegrate})
and system Jacobian calculations. This method was simple to implement, and familiar to those with a background
in C programming, but caused complications when combining BOUT++ with other libraries and frameworks.
In addition to this original style, an object-oriented style is now supported: Rather than callback functions,
users implement a class which inherits from \texttt{PhysicsModel}, overriding the default functions as needed. A
simple example is a diffusion equation in 1-D, which could be implemented in the following code:

\begin{verbatim}

class Diffusion : public PhysicsModel {
  private:
    Field3D T;
  protected:
    void init(bool restarting) { 
      SOLVE_FOR(T); 
    }
    void rhs(double time) { 
      ddt(T) = Laplace_par(T);
    }
};

\end{verbatim}

This defines a 3D scalar field $T$; specifies that $T$ should be evolved in the initialisation function \texttt{init};
and then calculates the time-derivative as $\frac{\partial T}{\partial t} = \nabla\cdot\left(\mathbf{b}_0\mathbf{b}_0\cdot\nabla T\right)$ in the function \texttt{rhs}. Users can use a macro \texttt{BOUTMAIN} to define a standard \texttt{main()} function, or define their own to enable BOUT++ to be combined with other libraries. Separating physics models into classes allows the possibility of combining several models into a single simulation, for example a model for neutral gas with a plasma model, and could be exploited for multiscale simulations.

The above example illustrates some other minor improvements which
have been made to BOUT++: \texttt{ddt()} and \texttt{SOLVE\_FOR} are preprocessor macros, which are used sparingly
wherever the resulting improvement in readability outweighs their potential for causing hard-to-find bugs.

\section{Solvers and capabilities}
\label{sec:capabilities}
 
As increasingly sophisticated plasma simulation models are studied with BOUT++, in particular gyro-fluid extensions~\cite{xu2013}, the range of differential operators which needs to be solved has expanded, and the computational cost of the simulations has increased. To address this, new capabilities have been added to BOUT++: More general elliptic solvers for calculating the electrostatic potential $\phi$ are presented in section~\ref{sec:boundaryvalue}, and an algorithm to solve parabolic equations along magnetic field lines is presented in section~\ref{sec:parabolicsolver}. This latter solver has been used as part of a physics-based preconditioning strategy to improve convergence for large time steps, described in section~\ref{sec:timeintegrate}, and to calculate closure terms for gyro-fluid operators~\cite{dimits2013}.
Wherever possible, these new solvers have been implemented using the factory pattern (section~\ref{sec:factory}) and a generic interface, so that they can be reused in many plasma models and geometries.

\subsection{Calculation of potential $\phi$ from vorticity}
\label{sec:boundaryvalue}

Reduced MHD models solved in BOUT++ are commonly formulated in terms of a
vorticity equation, from which the electrostatic potential is calculated. 
In reduced MHD, this can be derived from either the momentum equation or
charge conservation (current continuity)~\cite{hazeltine-2003,catto-2004}. Gyro-fluid models, which are an
area of current research in BOUT++~\cite{xu2013}, can also be cast in a vorticity
formulation~\cite{ottaviani1999}, or the potential can be calculated from a polarisation
equation coupling electron and ion gyro-centre densities~\cite{beer1997,scott-2005}. In either case,
the electrostatic potential $\phi$ is calculated by solving an equation
of the form:
\begin{equation}
\nabla\cdot\left(\frac{m_in}{B^2}\nabla_\perp \phi\right) = \omega
\label{eq:vorticity}
\end{equation}
with ion mass $m_i$, plasma density $n$, and magnetic field strength $B$. The time evolution of the right hand side $\omega$ (vorticity) depends on the particular model. The coefficient $n/B^2$ arises from the ion polarisation, and in general will vary in 3D as the density $n$ is an evolving quantity.  

Elliptic equations of the same form as equation~\ref{eq:vorticity} arise in many fields, and so numerical methods for their solution have been extensively studied. There are therefore many different methods available in the literature (e.g.~\cite{iserles2009}); the challenge is in finding one which is efficient enough for practical applications. 
Solving for $\phi$ requires the solution of a linear (matrix) problem for every evaluation of the time-derivatives of the system, which will usually be several times per time step.
A typical turbulence or ELM simulation might require $10^4 - 10^6$ time steps, and so
efficient solution to equation~\ref{eq:vorticity} is critical for the overall run-time
of the simulation.

A common approximation in plasma simulations is to neglect the variation
of $n/B^2$ in space and/or time, referred to as the Boussinesq approximation~\cite{yu2006}. 
BOUT++ simulations have usually replaced the full density $n$ in equation~\ref{eq:vorticity}
with the axisymmetric (constant in toroidal angle) equilibrium density $n_0$. Since $B$ is also axisymmetric, the left hand side of equation~\ref{eq:vorticity} can then be Fourier transformed in toroidal angle, decoupling the toroidal harmonics. Each toroidal harmonic can then be solved efficiently as a 1D tridiagonal system of complex equations in the radial coordinate. This scheme will be referred to here as the FFT or Boussinesq method. 

The effect of the Boussinesq approximation on simulation results can be subtle, and its importance depends on the problem being considered. As stressed elsewhere~\cite{scott-2002,scott03,scott2005-arxiv} the energetics of a model are important for long time simulations of turbulence. Unless the Boussinesq approximation is introduced carefully, it can result in an unphysical source of energy which can grow and eventually dominate the simulation. In other cases, the Boussinesq approximation has been found to have little effect, for example in BOUT++ simulations of blobs~\cite{angus2014} it was found that the Boussinesq approximation made only a small difference, and that in 3D the drift-wave dynamics made a greater difference to the result. 

To remove the Boussinesq approximation, and allow the solution to the full vorticity
equation, BOUT++ has been coupled to the PETSc library~\cite{efficient,petsc-user-ref}.
Here we present details of the numerical scheme,
and leave exploration of the impact on simulations of plasma phenomena to 
a future publication.

\subsubsection{Iterative solution with PETSc}

By discretising equation~\ref{eq:vorticity}, 
the calculation of $\phi$ from $\omega$ can be cast as a linear algebra problem of the form
\begin{equation}
\Mat{A}x = b
\end{equation}
In BOUT++ this discretisation is done by Finite Differences, but other choices
such as Finite Element are used elsewhere. The resulting matrix
can then be solved using the Krylov subspace (KSP) iterative solvers available in PETSc, 
such as GMRES. Iterative methods are attractive because they have smaller memory
requirements than direct solvers, as $\Mat{A}$ need never be explicitly stored,
and in principle iterative methods can be parallelised more efficiently.

In general, equation~\ref{eq:vorticity} will couple all points in the domain, so the $N\times N$ sparse matrix has a size $N\simeq 10^6$. By neglecting derivatives parallel to the magnetic field, which are assumed small in drift-ordered fluid~\cite{hazeltine-2003} and gyro-fluid~\cite{ottaviani1999} models, this can be simplified to solving multiple independent $N\simeq 10^4$ matrices. Since the density $n$ is evolving in time, the coefficients in this matrix change every time step. This makes direct solution
methods based on matrix factorisation inefficient, as the matrix must be frequently re-calculated
and re-factored. It is for this reason that iterative methods have been implemented in BOUT++, as these do not require the explicit calculation of the matrix elements or costly matrix factorisations. 

\subsubsection{Preconditioning of iterative solver}
\label{sec:boussprecon}

When solving large and/or ill-conditioned problems, iterative solvers can fail to converge, or converge very slowly after a small number of iterations (referred to as stalling). 
To accelerate convergence, an 
approximate solver is often used to ``precondition'' the problem, improving the 
condition of the matrix which the iterative solver is inverting. 
A preconditioner $\Mat{P}$ is an approximate inverse of $\Mat{A}$, which needs to be calculated quickly for the overall scheme to be efficient. The equation above can be multiplied through by $\Mat{P}$ as a {\it left preconditioner}:
\begin{equation}
\left(\Mat{P}\Mat{A}\right)x = \Mat{P}b
\end{equation}
or as a right preconditioner:
\begin{equation}
\left(\Mat{A}\Mat{P}\right)\left(\Mat{P}^{-1}x\right) = b
\end{equation}
and this modified system is solved using the iterative method. 
In the limit that the preconditioner $\Mat{P}$ is the inverse of $\Mat{A}$,
$\Mat{P}\Mat{A}$ is the identity, and no iterations should be required. 

To test preconditioning methods, a 2D ($x$, $z$) test case was used
with a density profile of the form:
\begin{equation}
n=\sin\left(x\right)e^{-x^2}\left(1 + p\cos\left(z\right)\right)
\end{equation}
where the radial coordinate $x$ goes between $0$ and $1$, and the constant $p$ in the above equation
is adjusted to change the variation of density with toroidal angle $z$. When
$p = 0$ density does not vary with toroidal angle, and so the Boussinesq approximation
is exact, but as $p$ is increased the approximation will break down. Results are shown
in table~\ref{tab:smalltiming} for a small $40\times 32$ mesh.
For three values of $p$ the time taken in seconds is given, followed by the iteration count in brackets
for a single solve, as the perturbation size is increased from 10\% to 90\% (i.e. $p$ varies from $0.1$ to $0.9$).
\begin{table}[h]
\centering
\caption{Timing for solution on a $40\times 32$ mesh. Shown are the wall clock times in seconds, and the iteration counts in brackets}
\label{tab:smalltiming}
\begin{tabular}[]{ccccc}
\hline
\hline
Preconditioner & \multicolumn{3}{c}{Density perturbation}\\
  & $10$\% & $50$\% & $90$\% \\
\hline
None & 0.157 (319) & 0.110 (226) &  0.142 (299) \\
\hline
Jacobi & 0.162 (318) & 0.113 (224) & 0.137 (299) \\
\hline
SOR & 0.022 (30) & 0.048 (35) & 0.048 (40) \\
\hline
\hline
FFT & 0.013 (4) & 0.015 (5) & 0.024 (9) \\
\hline
\hline
\end{tabular}
\end{table}
Without preconditioning the iterative method requires $\simeq 300$ iterations and $\simeq 1.5$ seconds to converge, compared to
a time of $\simeq 2$ms for a single Boussinesq solve. To improve on this, several ``black box'' preconditioning schemes
are available in PETSc, such as Jacobi iteration or Successive Over Relaxation (SOR) methods~\cite{iserles2009}. These methods are not
problem specific, and so require only a run-time switch to enable and configure. 
For the problems tested, the SOR method reduced the run-time (see table~\ref{tab:smalltiming}),
but the number of iterations remained prohibitive. To improve on these, a problem-specific preconditioner
has been implemented.

As discussed above, the purpose of a preconditioner is to quickly find an approximate
inverse to the linear problem (matrix $\Mat{A}$ above). The Boussinesq FFT-based solver is just such a
solver, as it finds a fast solution by simplifying the coefficients. We therefore
use the FFT solver as a preconditioner for the full problem,
by wrapping the FFT solver in a PETSc PCShell preconditioner object which is then 
passed to PETSc to be used in
the iterative solver. This preconditioner is extremely good when the density perturbation is small,
but we should expect it to become less effective as the size of the density perturbation becomes
large. This is what is observed in Table~\ref{tab:smalltiming}: For small perturbations (10\%), the run-time
is a little over half that of the SOR method, but as the density perturbation amplitude increases so does
the iteration count and run-time. Even at 90\% density perturbation, however, this preconditioner is still 
highly effective. Based on this small test, a larger study was performed to compare the SOR and FFT (Boussinesq)
preconditioners.

\begin{figure}[htbp!]
\centering
  \includegraphics[width=\columnwidth]{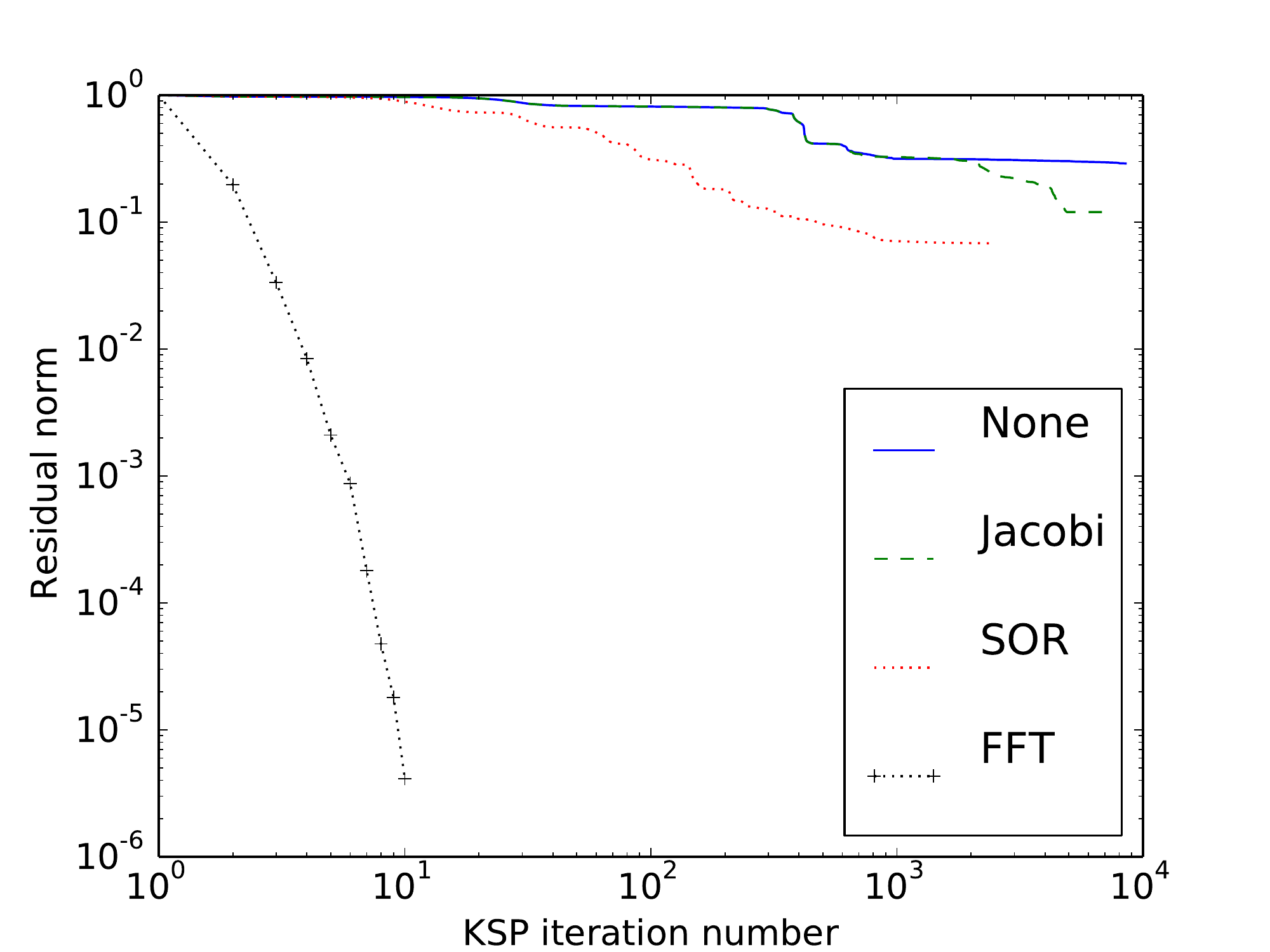}
\caption{Iterative KSP solver residual, as a function of iteration number. Solving elliptic equation~\ref{eq:vorticity} on a $516\times 256$ mesh.}
\label{fig:largemesh_precon}
\end{figure}

For a $516\times 256$ mesh more typical of ELM and turbulence calculations, the iterative solver will typically stall without good preconditioning, as shown in figure~\ref{fig:largemesh_precon}. Using a 90\% density perturbation (reducing the accuracy of the FFT preconditioner), and using the FFT solver result as the starting point for the iterative solver gives the convergence shown in figure~\ref{fig:largemesh_precon}.
Without preconditioning, or using Jacobi or SOR preconditioners, the residual is only reduced by a factor of $10$ in nearly $10^4$ iterations; using the FFT-based preconditioner the residual is reduced by a factor of $\simeq 10^5$ in $10$ iterations. This shows that for large meshes the FFT-based solver provides a good rate of convergence for realistic problem sizes, even when the deviation from axisymmetry is large. 

\subsection{Parabolic solver along magnetic field-lines}
\label{sec:parabolicsolver}

In the reduced MHD and gyro-fluid models which BOUT++ specialises in solving, the magnetosonic
fast wave is removed analytically, and so the fastest physical processes are usually the shear Alfv\'en wave
and heat conduction along (parallel to) magnetic field-lines. Preconditioning of either
of these processes for implicit time integration (see section~\ref{sec:timeintegrate}) requires the solution to a parabolic equation of the form
\begin{equation}
\left(\mathcal{A} + \mathcal{B}\partial_{||0}^2\right)x = b
\label{eq:parabolicPar}
\end{equation}
where $\partial_{||0} = \mathbf{b}_0\cdot\nabla$ is the derivative along the equilibrium magnetic field $\mathbf{b}_0$.
Even though equation~\ref{eq:parabolicPar} appears to be a one-dimensional problem, due to the structure of the equilibrium magnetic field, it is in general a two-dimensional problem: The equilibrium magnetic field in a tokamak is helical, and lies on nested toroidal surfaces. If the pitch angle of the magnetic field is such that it makes an irrational number of poloidal to toroidal turns, then a single field-line will fill the 2D surface. 

If magnetic perturbations were included, so that $\partial_{||0}^2$ became $\partial_{||}^2=\left(\mathbf{b}\cdot\nabla\right)^2$ in equation~\ref{eq:parabolicPar}, then the magnetic field lines in general no longer lie on magnetic flux surfaces, but fill a volume. This case is of significant interest, for example in studying the transport of heat in ELM crashes and in the presence of externally applied magnetic perturbations, but solving this more general problem is left to future work.

To solve equations of the form (\ref{eq:parabolicPar}), 
a solver has been implemented in BOUT++ using a
variant on the Thomas algorithm with interface equations \cite{austin2004}. In the following we assume
that the magnetic flux surface has $N$ points in toroidal angle, and $M$ points in poloidal angle. 
Firstly we exploit the toroidal symmetry of the equilibrium to decompose the problem into Fourier harmonics in
toroidal angle $\phi$. This then decomposes the problem into $N$ complex tridiagonal systems for each toroidal mode, each of size $M$. If the domain is a closed magnetic surface, then the tridiagonal systems are cyclic, with a complex phase shift between the first and last row which is determined by the pitch of the magnetic field-lines. Each
of these $N$ systems of equations therefore has the form:
\begin{equation}
\left(
\begin{array}{ccccccc}
b_0    & c_0    &         &         &        & a_0     &        \\
a_1    & b_1    & c_1     &         &        &         &        \\
       & \ddots & \ddots & \ddots  &        &         &        \\
       &        & a_{m-1} & b_{m-1}  & c_{m-1} &         &         \\ \hline
       &        &        & a_m     & b_m    & c_m      &        \\
       &        &        &         & \ddots & \ddots  & \ddots  \\
c_{M-1} &        &        &         &        &  a_{M-1} & b_{M-1}
\end{array}
\right) x = b
\label{eq:partridag}
\end{equation}
The domain is split between processors in the poloidal direction $\theta$, with $m$ rows per processor, illustrated by a
horizontal line in the above equation. Within each processor the rows are eliminated, reducing the problem to two
boundary equations for each processor, which forms a smaller (cyclic) tridiagonal matrix:
\begin{equation}
\left(
\begin{array}{cccc}
\beta_0  & \gamma_0 &          & \alpha_0 \\
\alpha_1 & \beta_1 & \gamma_1 &          \\ \hline
         & \alpha_2 & \beta_2 & \gamma_2 \\
\gamma_3 &          & \alpha_3 & \gamma_3
\end{array}
\right) \chi = \xi
\end{equation}
where $\xi$ and $\chi$ are the boundary values of $b$ and $x$ respectively. 
When solving $N$ independent systems of equations (one for each toroidal mode), they are
divided between processors, and all boundary equations for a given 
system are gathered onto a single processor. For example if $N$ systems are split between 2 processors,
then $N/2$ sets of boundary equations are gathered onto each processor. Once on a single processor,
the serial Thomas algorithm (with Shermann-Morrison formula for cyclic tridiagonal systems) is used to solve
for the boundary values. These are then scattered back, and substituted into the original equation
to obtain the solution inside each processor's domain. 

\begin{figure}[htbp!]
\centering
  \includegraphics[width=\columnwidth]{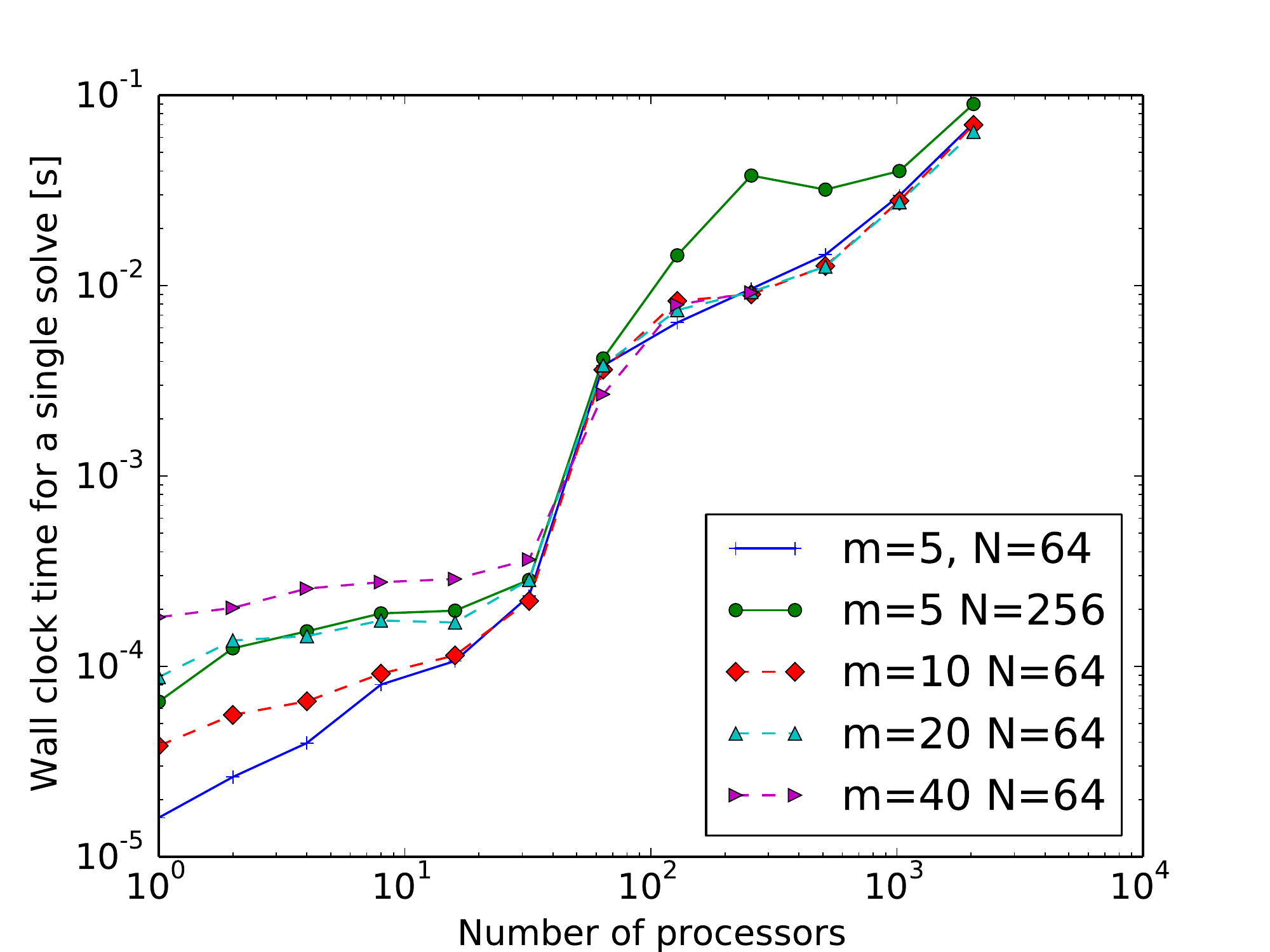}
\caption{Soft scaling of parallel solver. $m$ is the number of rows per processor, and $N$ is the number of separate problems which are inverted simultaneously, corresponding to the number of toroidal Fourier modes}
\label{fig:scaling_cyclic}
\end{figure}

The number of boundary equations for each system of equations, and number of communications
is independent of the size of the problem $M$, and proportional to the number of processors. 
This means that the algorithm scales well with problem size, but uses gather and scatter operations 
which reduces performance on large numbers of processors. Scaling of the solver with processor
number and problem size has been performed on HECToR, with 32 cores per node, and up to 2048 cores in total.
Soft scaling, in which the problem size is increased proportionally with the number of processors,
is shown in figure~\ref{fig:scaling_cyclic}.
The factor of $\simeq 10$ increase in wall clock time above 32 processors in figure~\ref{fig:scaling_cyclic}
is because above this point communications occur between nodes, rather than solely within a single node. 
Waiting for global gather operations from across nodes is significantly slower than within a node,
and so becomes a bottleneck: for more than 64 processors the run time becomes almost independent of 
problem size $m$.

\begin{figure}[htbp!]
\centering
  \includegraphics[width=\columnwidth]{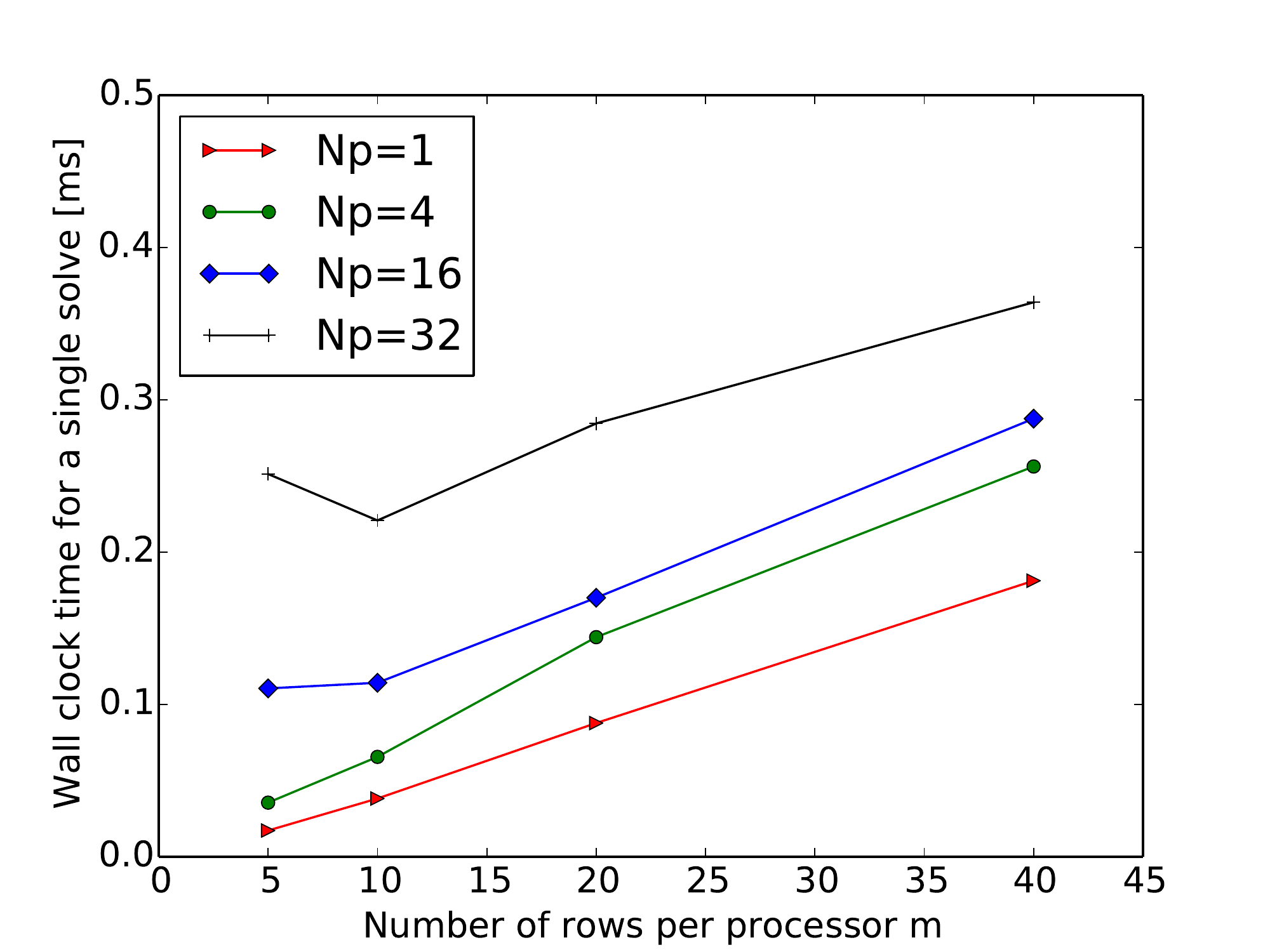}
\caption{Soft scaling of parallel solver showing same data as figure~\ref{fig:scaling_cyclic} for $N=64$. 
Shows variation of wall-clock time with problem size per processor $m$ for different numbers of processors $N_p$.}
\label{fig:scaling_cyclic2}
\end{figure}

For fewer than 32 processors, it can be seen that doubling the number of independent systems
$N$ has a similar effect to doubling the size of each system $M$, and the wall time
is approximately proportional to the problem size. This can also be seen in figure~\ref{fig:scaling_cyclic2}:
on single core the algorithm scales approximately linearly with problem size, for larger numbers of 
processors there is a constant offset which depends on the number of processors, but only weakly
on the number of rows per processor.
The result is that as the number of rows per processor $m$ is increased the scaling with processor number
appears more favourable: In the limit that each processor only has a two rows, the algorithm becomes equivalent
to gathering all rows onto one processor, and using a serial algorithm, which would result in a linear scaling
with number of processors $N_p$. In figure~\ref{fig:scaling_cyclic}, when $m = 5$, the wall time scales with
$N_p^{0.77}$, whilst for $m = 40$ the exponent is $\simeq 0.20$. 

This scaling analysis demonstrates that the use of boundary equations to reduce the problem within
each domain to two rows, results in an approximately linear scaling with problem size on a fixed number of processors.
The global gather and scatter operations used to solve these boundary equations is a bottleneck for large numbers
of processors, and leads to poor parallel scaling. Improving this will be the subject of future investigation. A possible solution is to use the cyclic reduction algorithm to solve the boundary value equations, which would eliminate global gather/scatter operations in favour of more point-to-point communications. In practice, the current solver
has been found to be sufficiently fast for current simulations: the domain is decomposed in both radial and poloidal
directions, so the poloidal direction is typically not divided into more than 32 processors. In addition,
the total run-time spent in solving parallel parabolic equations is a small fraction of the total. As
a result, this method has enabled the use of algorithms which have overall good scaling~\cite{dudson2012,dimits2013}.

This tridiagonal solver has been used to implement physics-based preconditioning in BOUT++~\cite{dudson2012},
which is discussed in section~\ref{sec:timeintegrate}. It has also been used to implement 
gyro-fluid parallel closures~\cite{dimits2013,xu2013} approximating to high accuracy the $k / \left|k\right|$
operators appearing in closures such as those due to Hammett-Perkins~\cite{hammett90} by a small number of Lorentzians,
which take the form of equation~\ref{eq:parabolicPar}. This is discussed further in section~\ref{sec:nonlocal}.

\subsection{Time integration}
\label{sec:timeintegrate}

The time integration schemes available in BOUT++ has been expanded to
include explicit schemes (RK4, Karniadakis~\cite{karniadakis1991}, Euler), and a
range of implicit schemes through coupling to the SUNDIALS~\cite{hindmarsh2005} and PETSc~\cite{efficient,petsc-user-ref} libraries. 

The need to solve increasingly complicated models at increasingly high resolution has made
efficiency and parallel scaling of algorithms used important. Plasma models are typically stiff,
meaning that explicit time-integration methods are limited to small time-steps relative to the time-scales 
of interest. Implicit time
integration methods overcome this restriction, but require the solution of a large linear system
($N\times N$ where $N$ is the number of evolving variables, typically $1-10$ million) at every timestep. Unless this can be done
efficiently, the overall time taken by an implicit solver may not be less than an explicit solver. 

As discussed in section~\ref{sec:boussprecon}, a preconditioner is an approximate inverse of the large matrix being solved, which should be efficient to
evaluate. At each timestep an implicit time integration method is solving a nonlinear problem to find the state at the next time point. Using a scheme such as Newton's method, this nonlinear problem is reduced to one or more steps which require the solution to a linear problem of the form $\mathcal{A}\mathbf{x} = \mathbf{b}$
where $\mathbf{b}$ is known, and depends on current and previous state and their time-derivatives; $\mathbf{x}$
is related to the unknown state at the next timestep; and $\mathcal{A}$ is a large matrix. Advancing a single time-step therefore involves an outer loop to solve the nonlinear
problem, which contains an inner loop to find the solution to a series of linear problems. 
Preconditioning targets this inner loop, improving convergence of the linear solve in order to reduce the overall run time.

Physics-based preconditioning describes a family of approaches to deriving preconditioners,
which use knowledge of the physical system to simplify the model equations. The aim is to retain in the preconditioner only those processes (oscillatory or diffusive) which are limiting the timestep. This then improves the condition number of the matrix which the iterative (usually Krylov subspace) method has to solve. Because the iterative method converges towards the solution to the full system, approximations can be made in the preconditioner without affecting the final solution, only the convergence rate towards the solution. The approach we have followed is based on work by Chacon and others~\cite{ChaconKnollEtAl02}. Preconditioning of the implicit solvers in BOUT++ (using SUNDIALS and PETSc libraries) has been implemented in BOUT++, described in~\cite{dudson2012}. 

An example is the shear Alfv\'en wave which is present in all 3D turbulence models. 
In the simplest form of the reduced MHD equations, this wave can be described by
the following coupled equations for vorticity $\omega$ (equation~\ref{eq:vorticity}),
and magnetic potential $A_{||}$ which describes the perturbed magnetic field $\delta\mathbf{B} = \nabla\times\left(\mathbf{b}_0 A_{||}\right)$:
\begin{eqnarray}
\frac{\partial\omega}{\partial t} = \nabla_{||0}\left(j_{||}\right) \qquad \frac{\partial A_{||}}{\partial t} = -\partial_{||0}\phi \label{eq:alfvenwave}\\
\omega = \nabla\cdot\left(\frac{m_in}{B^2}\nabla_\perp\phi\right) \qquad \nabla_\perp^2 A_{||} = -\mu_0j_{||} \nonumber
\end{eqnarray}
where $\nabla_{||0}f = \nabla\cdot\left(\mathbf{b}_0f\right)$, $\partial_{||0} = \mathbf{b}_0\cdot\nabla$, $\nabla_\perp f = \nabla f - \mathbf{b}_0\left(\mathbf{b}_0\cdot\nabla f\right)$, and $\nabla_\perp^2 f = \nabla\cdot\left(\nabla_\perp f\right)$. 
These equations can be combined to give a wave equation, which in the case that $n/B^2$ is a constant reduces to:
\begin{equation}
\frac{\partial^2\omega}{\partial t^2} = \nabla_{||0}\frac{\partial j_{||}}{\partial t} = \nabla_{||0}\left(\frac{1}{\mu_0}\nabla_\perp^2\partial_{||0}\frac{B^2}{m_in}\nabla_\perp^{-2}\omega\right)
\end{equation}
By further assuming that the magnetic field $\mathbf{B}$ varies slowly, and so neglecting derivative of $B$ terms,
\begin{equation}
\nabla_{||0}\nabla_\perp^2\partial_{||0}\nabla_\perp^{-2} \simeq \partial_{||0}^2
\end{equation}
and the shear Alfv\'en wave propagates only along the (equilibrium) magnetic field:
\begin{equation}
\frac{\partial^2\omega}{\partial t^2} = \frac{B^2}{\mu_0m_in}\partial_{||0}^2\omega = V_A^2\partial_{||0}^2\omega
\label{eq:alfvenwaveeq}
\end{equation}
where $V_A$ is the Alfv\'en speed. In tokamak simulations, this speed can be $V_A\simeq 10^7$m/s, severely restricting the time step for explicit time integration schemes. 
If hyperbolic equation~\ref{eq:alfvenwaveeq}, or the original equations~\ref{eq:alfvenwave} are solved
implicitly, then this requires the solution to a parabolic equation. For example using a backwards Euler method:
\begin{equation}
\left(\begin{array}{c}
\omega \\
\omega'
\end{array}
\right)^{t+1}
 = \left(\begin{array}{c}
\omega \\
\omega'
\end{array}
\right)^t + \delta t \left(\begin{array}{cc}
0 & 1 \\
V_A^2\partial_{||0}^2 & 0
\end{array}
\right) \left(\begin{array}{c}
\omega \\
\omega'
\end{array}
\right)^{t+1}
\end{equation}
the equation to be solved at each timestep is parabolic:
\begin{equation}
\left(1 - \delta t^2 V_A^2\partial_{||0}^2\right)\left(\begin{array}{c}
\omega \\
\omega'
\end{array}\right)^{t+1} = \left(\begin{array}{cc}
1 & \delta t \\
\delta tV_A^2\partial_{||0}^2 & 1
\end{array}
\right)\left(\begin{array}{c}
\omega \\
\omega'
\end{array}\right)^t
\label{eq:parabolic}
\end{equation}
and of the same form as equation~\ref{eq:parabolicPar}. To precondition waves of this type, the parabolic solver discussed in section~\ref{sec:parabolicsolver} can therefore be used.

The assumptions made to reduce the full set of equations~\ref{eq:alfvenwave} to wave equation~\ref{eq:alfvenwaveeq} cannot be made in the calculation of time-derivatives for the full problem, as this would affect the solution, but they can be made in the preconditioner since this is used to find an approximate solution to accelerate convergence to the solution of the full set of equations. The full procedure to derive a preconditioner using Schur factorisation is described in~\cite{dudson2012}, and is somewhat more involved than outlined here, but it makes the same assumptions and so results in the same form of equations. The resulting preconditioner using the solver presented in section~\ref{sec:parabolicsolver} has been found to result in significant speed-ups for sets of equations where the timestep was limited by the shear Alfven wave and parallel heat conductivity~\cite{dudson2012}, reducing overall wall-clock time by an order of magnitude in some cases.

\subsection{Non-local heat transport}
\label{sec:nonlocal}

Heat transport along magnetic field-lines plays a crucial role in determining the flux of heat to material surfaces in tokamak devices, and so is of importance to the design of next-generation machines and demonstration power-plants. In the collisional limit, the heat flux is described by the well-known Spitzer formula~\cite{wesson-1997}, but at the high temperatures relevant to the edge of large tokamaks, the mean-free-path of electrons along the magnetic field can become comparable to the system size: Typical values in JET of $T_e\simeq 100$eV and $n_e\simeq 10^{19}$m$^{-3}$ give an electron mean free path of $\lambda\simeq 14$m, whilst the connection length from midplane to outer divertor target is of the order of $30$m.  In these situations flux limiters are often employed~\cite{schneider2006}, which reduce heat flux to the free-streaming limit. Unfortunately these methods often perform poorly when compared to kinetic solutions~\cite{CTPP:CTPP201210038,CTPP:CTPP200810015}, and contain a free parameter which must be determined. There is therefore interest in developing first-principles heat flux models which better approximate the kinetic solutions whilst minimising the computational cost. 

Two such methods have been implemented in BOUT++: an extension of the Hammett-Perkins model to non-Fourier methods~\cite{dimits2013}, and a method based on solving a 1-D time-independent kinetic equation along magnetic field-lines~\cite{ji:022312}, which has been shown to reproduce the collisional and collisionless limits, and applied to ELM simulations \cite{omotani2013}.

\section{Pre- and Post-processing tools}
\label{sec:tools}

A simulation code is of little use without the tools to create high quality inputs such as meshes,
and to analyse and present the results of the simulations. For post-processing the most important
requirement is to be able to read the simulation output data in the user's language of choice. 
Routines to do this are now available for IDL, Python, Matlab, Octave, and Mathematica as part
of the public BOUT++ repository. 

For most publications, 1D plots and 2D contours are sufficient, but there
are occasions when the ability to visualise data in three dimensions is
useful. In the early stages of a scientific investigation, seeing the entire
simulation domain rather than slices through it, can help spot anomalies or
unexpected features. When presenting results, particularly to conferences,
3-D images visualisations can quickly convey a large amount of information.
Wrappers have been developed to enable two scientific data visualisation
packages to be used with BOUT++: Mayavi\footnote{Mayavi project, \link{http://code.enthought.com/projects/mayavi/}} and VisIT\footnote{VisIT tool \link{https://wci.llnl.gov/codes/visit}}.

Tools have also been developed to enable more convenient input of initial profiles and sources (section~\ref{sec:inputexpr}), and processing of experimental equilibria into input meshes (section~\ref{sec:meshgen}).

\subsection{Input expression parser}
\label{sec:inputexpr}

\begin{figure}[htbp!]
\centering
  \includegraphics[]{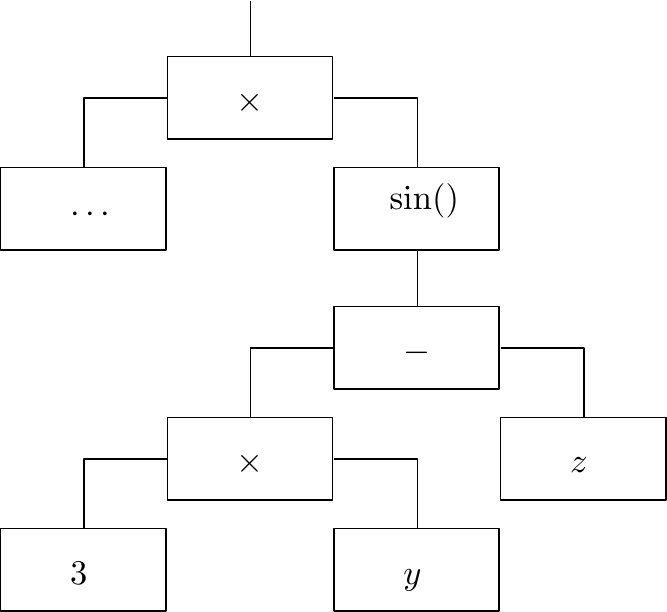}
\caption{A tree of generator objects (Abstract Syntax Tree) to evaluate the expression \texttt{(...)*sin(3*y-z)}}
\label{fig:ast}
\end{figure}

Many simulations do not require complex geometry, but are intended to study basic physical mechanisms in slab
or cylindrical geometries. 
For these cases the initial conditions and parameters often follow analytic expressions. If these expressions can be stored in the input files rather than preprocessing scripts, then inputs can be modified more quickly, and a clearer record of simulation inputs is kept for later reference. Examples include 2D blob simulations, where the initial density profile is a Gaussian in $x$ specified using:

\begin{verbatim}

function = 1 + 0.2*gauss(x-0.25, 0.1)

\end{verbatim}

In slab simulations of forced reconnection,
a helical external magnetic potential is applied to an initial sheared magnetic field. 
This external field can be specified in the input file using:

\begin{verbatim}

function = (1-4*x*(1-x))*sin(3*y - z)

\end{verbatim}

A recursive descent parser with operator precedence (see e.g.\cite{compilers,llvmtutorial}) is used to build an Abstract Syntax Tree (AST) of generator objects from the input text: A constant generator like '3' or '\texttt{pi}' always returns the same value; a coordinate generator like '\texttt{x}' returns a value depending on the cell location; and a binary operation generator like '\texttt{+}' or '\texttt{sin}' depends on the value of its children generators. Part of the AST for the above example is shown in figure~\ref{fig:ast}.
This tree is then evaluated for each cell in the domain to obtain the required initial conditions. This method is not computationally efficient, but is only used during initialisation and provides the flexibility to adapt the code in future. 

Initially a learning exercise in compilers, the capability to construct ASTs from inputs, then manipulate and execute them at runtime has proved to be useful for scientific applications, eliminating the need to write input preprocessing scripts in many cases. This is
proving particularly useful for verification using the Method of Manufactured Solutions~\cite{roache1998,salari2000}, in which analytic expressions for sources and boundary conditions need to be specified. This verification activity is ongoing, and will be presented in a future publication.

As demonstrated by libraries such as SciPy~\cite{scipy}, the combination of an efficient statically compiled library with the flexibility of an run-time interpreter can be very powerful. 
A possibility for future exploration is to use a scripting language such as Python, Lua, or Scheme to
provide input and output to BOUT++, or even to implement parts of physics
modules. These languages provide a complete programming environment, but would require significant work to interface with the C++ classes in BOUT++ than the algebraic expression parser implemented currently.

\subsection{Mesh generation} 
\label{sec:meshgen}

The accuracy and robustness of plasma simulations is strongly dependent on the quality of the input mesh: noise in the metric components, or large variations in the grid spacing leads to noise in the solution, restrictions in the timestep, and occasionally numerical instabilities. 

Generating meshes for tokamak equilibria with X-points is challenging due to the change in topology across the separatrix, the shape of the boundary around the plasma, and the variation in geometry between machines.
The original BOUT code used UEDGE~\cite{rognlien1992,rognlien-2002} to generate grids, and BOUT++ can also use these input files with a little preprocessing. There are several other
tools available for generating these meshes such as CARRE~\cite{marchand1996}, but none are open-source licensed and suitable for distribution with BOUT++. 

To generate X-point meshes for BOUT++ from experimental free boundary equilibria, a new code named \noun{Hypnotoad} has been developed, and is available in the BOUT++ public repository.
The main features of this grid generator are that it:
\begin{enumerate}
\item Was originally written entirely IDL. This allows Hypnotoad to be run anywhere where IDL is available,
without the need for a compilation step and complicated dependencies. IDL is widely available and used in fusion
research institutions, and comes with a large library of built-in functions. Work is currently ongoing to port the algorithms which have been developed into Python, and indeed many of the figures shown here will be from the Python version, due to the superior graphical capabilities of the Python library Matplotlib\footnote{Matplotlib library \link{http://matplotlib.org/}}.
\item Automatically adjusts settings when needed. The grid produced can
    be customised, but the minimum number of inputs is very small
    (number of grid points and a range of poloidal flux $\psi$). The $\psi$ range asked for is
    adjusted to fit within the boundary, with configurable levels of strictness.
\item Can handle an arbitrary number of X-points. Whilst not of obvious
    benefit since most tokamak equilibria are single or double-null, this means that 
    \noun{Hypnotoad} is quite generalised and can cope with unusual configurations. It has been applied to Snowflake-like configurations~\cite{ma2014}, but only in snowflake-plus configurations where the second X-point was not included in the mesh.
\end{enumerate}
Because this grid generator is intended to be used for many different tokamaks, robust algorithms
have been developed which can handle the large number of possible configurations encountered. 
To date, this grid generator has been successfully used to generate meshes from
 C-MOD, DIII-D, EAST, ITER, JET, MAST and NSTX equilibria, without requiring any machine-specific alterations or inputs beyond the EFIT generated 'g' file. 

The production of a BOUT++ input mesh consists of three stages: Analysis of the equilibrium to determine X-point locations, construction of mesh points, and calculation of metric tensor and equilibrium quantities. The key features and algorithms used in each of these stages are described in the following sections.

\subsubsection{Finding X-points}

The first task in generating a mesh is to determine the number and location of the O- and X-points of the plasma. These correspond to critical points (maxima, minima, or saddle points) of the poloidal flux function $\psi\left(R,Z\right)$, which for tokamak equilibria is a 2D function of major radius $R$ and height $Z$. 
The most robust technique for finding these has been found empirically to be to produce contour lines of $\frac{d\psi}{dR} = 0$ and $\frac{d\psi}{dZ} = 0$. Intersections of these lines then give locations of critical points. By comparing second derivatives of $\psi$ at these critical points it can be determined whether they are O-points (minima/maxima) or X-points (saddle points). 
\begin{figure}[htbp!]
\centering
  \includegraphics[]{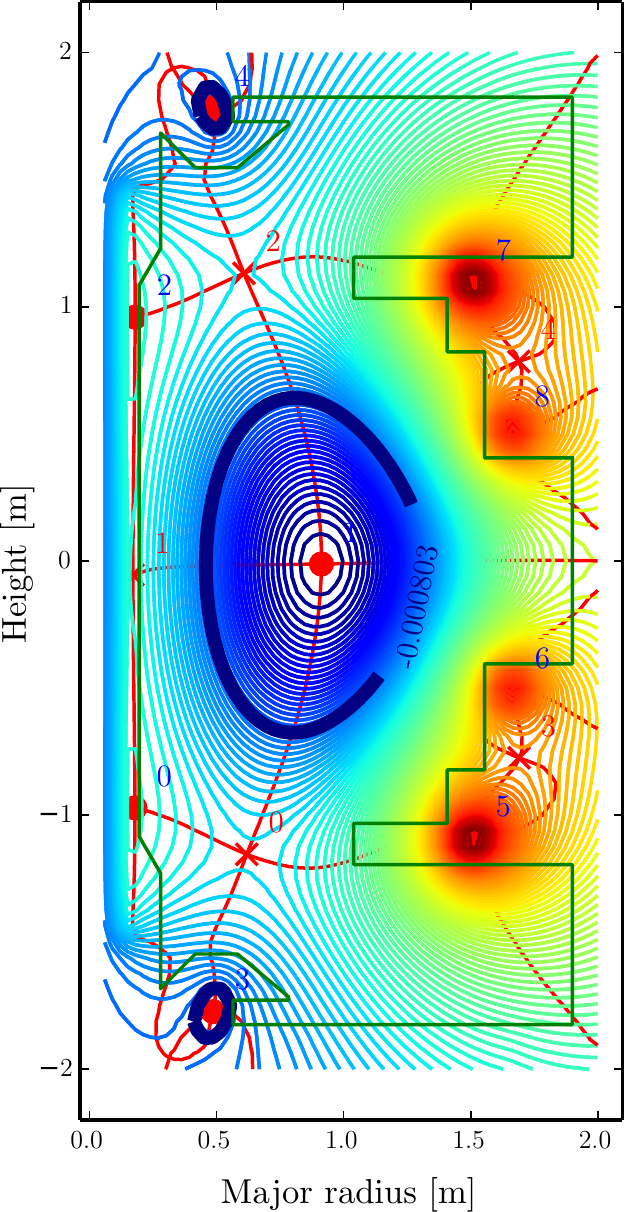}
\caption{Automated identification of O- and X-points in a MAST double-null configuration}
\label{fig:mast_grid_py}
\end{figure}
An example of a double-null equilibrium from the Mega-Amp Sherical Tokamak (MAST) is shown in figure~\ref{fig:mast_grid_py}. Contours of $\frac{d\psi}{dZ} = 0$ and $\frac{d\psi}{dR} = 0$ are plotted, and their intersections identified and categorised. 
   Some additional heuristics are needed to eliminate false positives or duplicate critical points, which can occur if the input data contains grid-scale features or noise. The primary O-point can be reliably identified as the one closest to the middle of the grid, but identifying the plasma X-points is more prone to error, as a typical equilibrium will contain several X-points. 

When the boundary shape is
specified, critical points close to the poloidal field coils can be discarded as being outside the boundary, though ripples in the solution due to the central solenoid at major radius $R=0$ can lead to spurious O- and X-points on the inboard side: In figure~\ref{fig:mast_grid_py} two O-points (labelled '0' and '2'), and one X-point (labelled '1') can be seen close to the centre column. These can usually be excluded through choice of $\psi$ range or boundary contour.

\subsubsection{Meshing}

\begin{figure}[htbp!]
\centering
  \includegraphics[width=\columnwidth]{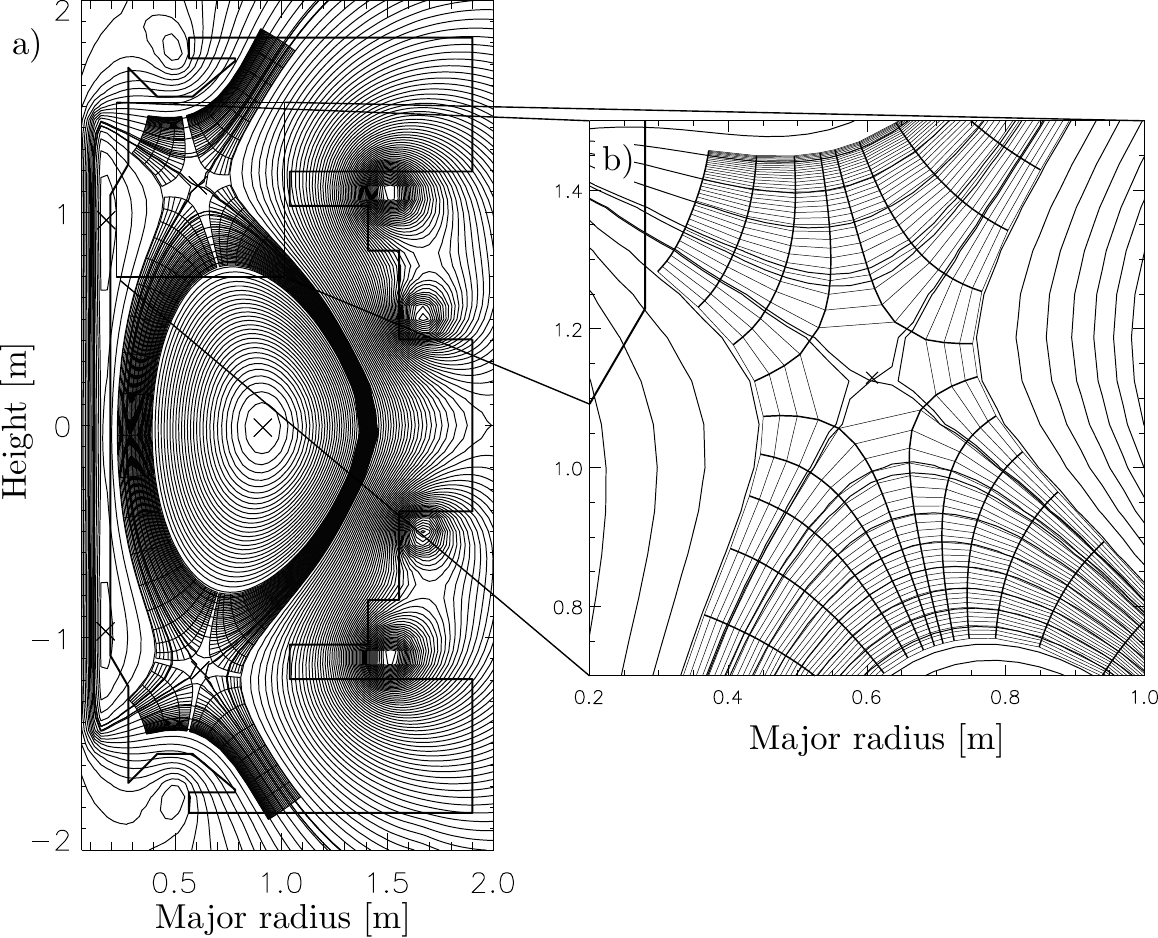}
\caption{Mesh produced for MAST double-null equilibrium (shot 14220)}
\label{fig:mast_mesh}
\end{figure}

Transport of heat and particles in magnetised plasmas is strongly anisotropic,
and as a result fluid simulations commonly use meshes aligned with
magnetic flux surfaces, in order to minimise mixing the directions perpendicular
and parallel to the magnetic field. The coordinates currently used by BOUT++ 
for tokamak simulations are orthogonal in the poloidal (R-Z) plane, illustrated in figure~\ref{fig:mast_mesh}
for a typical MAST double-null discharge. 
A region close to the X-point is shown
enlarged, showing coordinate lines passing around the X-point, but leaving a hole at the null itself.
The mesh has a branch-cut around the magnetic X-point, which must be treated
carefully to avoid creating large variations in mesh spacing, which could lead to
poor numerical behavior. 

A starting contour line is created, aligned with a magnetic flux surface. In the plasma core this line is just inside
the separatrix, whilst for the divertor legs the separatrix is used. From this starting
line the locations of the X-points are found, and the regions between X-points
are then meshed independently, adjusting the distance to the X-point to obtain
a smoothly varying mesh spacing. From these starting locations the gradient of
$\psi$ is found using either a Discrete Cosine transform (DCT) method, or 2D splines. The
DCT method provides smooth interpolation, but is slow for large input meshes and sharp
features in the input can lead to ringing (Gibbs phenomena). The spline method is therefore
generally preferred. The gradient in $\psi$ is followed using the LSODE algorithm through IDL or SciPy, to
construct the mesh points for each value of $\theta$. This is then repeated for each $\theta$
coordinate to construct a 2D mesh. Once the mesh points have been found, the metric tensor 
and equilibrium components are calculated.

\subsubsection{Calculating metric components}

Experimental equilibria are often of low resolution ($65\times 65$ is the standard EFIT output for many tokamaks),
and once the pressure $P$ and magnetic field $\mathbf{B}$ is interpolated onto a new mesh there is
no guarantee that the new values will still obey ideal MHD force balance
$\left(\nabla\times\mathbf{B}\right)\times\mathbf{B} = \mu_0\nabla P$. Ideal MHD force balance may not 
be an equilibrium solution to the plasma model being simulated, but is almost invariably a good approximation, 
and it is important that the mesh generation process does not lead to artefacts or sources of numerical noise and instability. Care is therefore taken to ensure the accuracy and smoothness of the interpolated solution, and quantities such as parallel current density $J_{||}$ are calculated multiple ways and compared as a consistency check.
The interpolation method used can have a significant impact on the quality of the results: 
many terms such as the curvature and parallel current involve second derivatives of equilibrium
flux, and these quantities should themselves be smoothly varying inputs
to the simulation. 

As discussed elsewhere~\cite{xu-2008,Dudson2009}, in order to efficiently simulate structures (predominantly) aligned to the magnetic field, 
BOUT++ uses grid-points placed in a field-aligned coordinate system.
From the standard, orthogonal, toroidal coordinate system $\left(\psi, \theta, \zeta\right)$ new coordinates $\left(x,y,z\right)$:
\begin{eqnarray}
x &=& \sbt\left(\psi - \psi_0\right) \qquad y = \theta \\
\label{eq:coordtransform}
 z &=& \sigma_{B\theta}\left(\zeta - \int_{\theta_0}^{\theta}\nu\left(\psi, \theta\right)d\theta\right) \nonumber
\end{eqnarray}
where $\sigma_{B\theta} \equiv B_\theta / \left|B_\theta\right|$ is the sign of the poloidal field, 
and $\nu$ is the local field-line pitch given by
\begin{equation}
\nu\left(\psi, \theta\right) = \frac{\mathbf{B}\cdot\nabla\zeta}{\mathbf{B}\cdot\nabla\theta} = \frac{\Bt h_\theta}{\Bp R}
\end{equation}
where $\Bt$ is the toroidal magnetic field, $\Bp$ the poloidal magnetic field, $R$ is the major radius, and $h_\theta$ is poloidal arc length divided by $2\pi$. In the limit of a circular cross-section equilibrium, $h_\theta$ becomes the minor radius $r$.
The coordinate system is chosen so that $x$ increases radially outwards, from plasma to the wall.
The sign of the toroidal field $\Bt$ can then be either positive or negative.

\begin{figure}[htbp!]
\centering
\subfigure[Radial pressure profiles from input (solid line) and from equation~\ref{eq:xbalance} for radial force balance (symbols)]{
  \label{fig:forcebalance_p}
  \includegraphics[width=0.9\columnwidth]{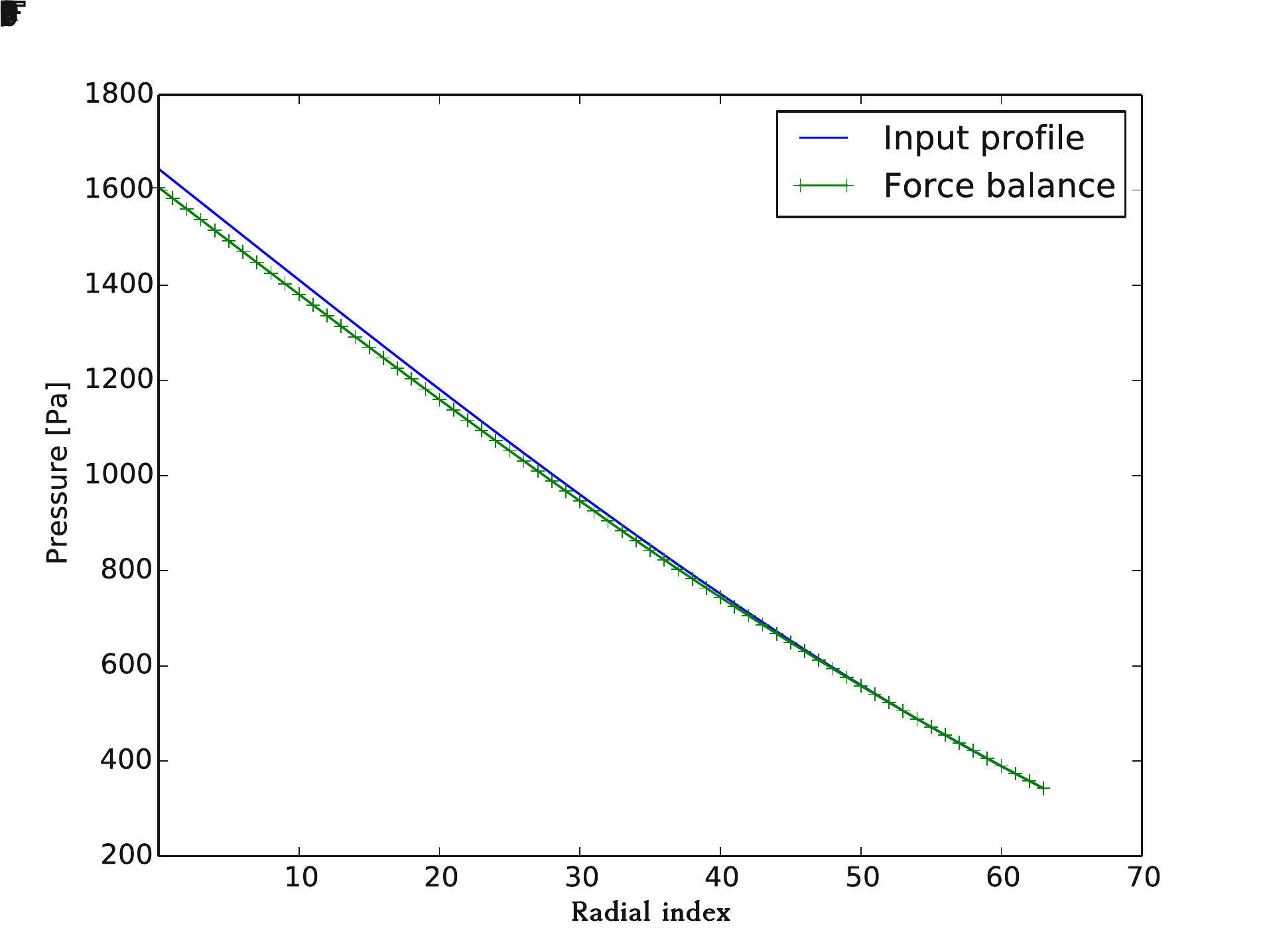}
}
\subfigure[Radial profile of poloidal arc length $h_\theta$ measured geometrically (solid line), and using equation~\ref{eq:xbalance} for radial force balance (symbols)]{
  \label{fig:forcebalance_hthe}
  \includegraphics[width=0.9\columnwidth]{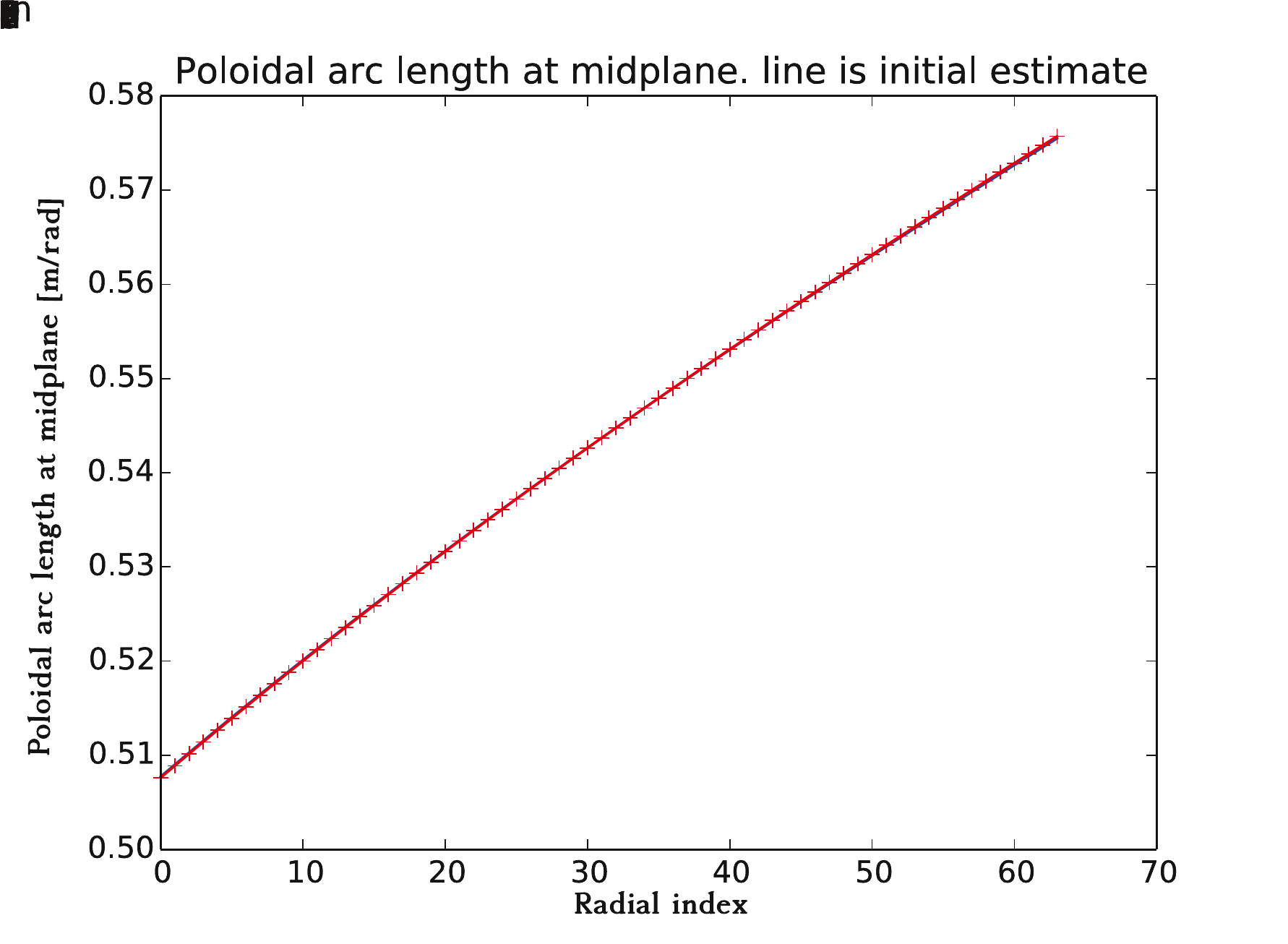}
}
\caption{Comparison of pressure and $h_\theta$ profiles calculated using radial force balance (equation~\ref{eq:xbalance}) with a reference calculated independently.}
\label{fig:forcebalance}
\end{figure}

By equating contravariant $x$ components of $\Jvec\times\Bvec = \nabla P$, radial force balance
in field-aligned coordinates can be written as:
\begin{equation}
\deriv{}{x}\left(\frac{B^2\hthe}{\Bp}\right) - \Bt R\deriv{}{x}\left(\frac{\Bt\hthe}{R\Bp}\right) + \frac{\mu_0\hthe}{\Bp}\deriv{P}{x} = 0
\end{equation}
Close to the X-points, $\Bp\rightarrow 0$ and the above expression becomes singular, so a better way to write this is:
\begin{eqnarray}
\deriv{}{x}\left(B^2\hthe\right) &-& \hthe\Bp\deriv{\Bp}{x} \nonumber \\
&-& \Bt R\deriv{}{x}\left(\frac{\Bt\hthe}{R}\right) + \mu_0\hthe\deriv{P}{x} = 0
\label{eq:xbalance}
\end{eqnarray}
This expression is used to calculate the pressure, by integrating $\deriv{P}{x}$, and compared with the input pressure profile. By using the input pressure profile for $P$, $h_\theta$ is also calculated and compared with the values calculated from geometric arc-length. An example result of this comparison is shown in figure~\ref{fig:forcebalance}.
In figure~\ref{fig:forcebalance_p} the pressure at the outermost radial point has been set to the value from the EFIT input, and then integrated inwards according to equation~\ref{eq:xbalance}. In figure ~\ref{fig:forcebalance_hthe} the pressure gradient from EFIT has been used, and equation~\ref{eq:xbalance} solved for $h_\theta$. Both show good agreement, indicating that the mesh generation is sufficiently accurate and smooth to retain radial force balance. This is not always observed, and where large discrepancies are observed the equilibrium should not be used.

Many reduced MHD models make use of the quantity 
\begin{equation}
\frac{B}{2}\nabla\times\left(\frac{\mathbf{b}}{B}\right) \simeq \mathbf{b}\times\mathbf{\kappa}
\end{equation}
where $\mathbf{\kappa} = \mathbf{b}\cdot\nabla\mathbf{b}$ is the curvature vector, 
which arises from magnetic particle drifts. Calculation of this quantity involves second derivatives of the input poloidal flux $\psi$, and so must be calculated carefully to avoid introducing noise. Several methods have been tried, including:
\begin{enumerate}
 \item Calculate curvature on the original R-Z mesh supplied
       as input, then interpolate onto the new field-aligned mesh. This has been found to usually produce the smoothest result and so is the default. Using DCTs to calculate differentials of the magnetic field components was found to produce oscillatory results, so a 3-point Lagrangian interpolation is used instead.
 \item Calculate curvature on the field-aligned mesh in toroidal coordinates, using nearest neighbours and least-square fitting.
 \item Calculate curvature in field-aligned coordinates using
\begin{eqnarray}
\nabla\times\left(\frac{\bvec}{B}\right) &=& \frac{\Bp}{\hthe}\left[\left(\deriv{}{x}\left(\frac{\hthe}{\Bp}\right) - \deriv{}{y}\left(\frac{\sbt\Bt IR}{B^2}\right)\right)\Vec{e}_z \right. \nonumber \\
&& + \deriv{}{y}\left(\frac{\sbt\Bt R}{B^2}\right)\Vec{e}_x \nonumber \\
&& + \left.\deriv{}{x}\left(\frac{\sbt\Bt R}{B^2}\right)\Vec{e}_y\right]
\end{eqnarray}
\end{enumerate}
Methods (ii) and (iii) are based on calculating differentials on the output (field-aligned) mesh. They work well when the input is of high resolution, but become noisy once the resolution of the output grid significantly exceeds that of the input. Since high resolutions are required for BOUT++ simulations, this is nearly always the case, and so method (i) is preferred. Methods (ii) and (iii) are retained for cross-comparison.

\begin{figure}[htbp!]
\centering
  \includegraphics[width=0.8\columnwidth]{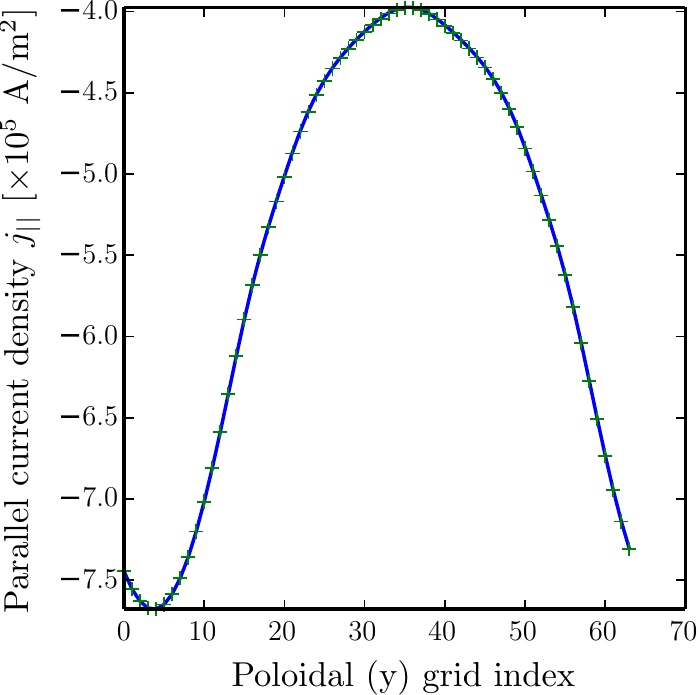}
\caption{Comparison of $j_{||}$ from input profiles (equation~~\ref{eq:inputcurrent}) and divergence of current (equation~\ref{eq:diveqj})} 
\label{fig:jparcomparison}
\end{figure}

Finally, a cross-check is made between the parallel
current and the curvature. In a tokamak equilibrium
the divergence of the parallel current balances the 
divergence of diamagnetic current, so that the total current is divergence free. In reduced MHD
models this appears through the vorticity equation. 
The following relationship should therefore be satisfied:
\begin{equation}
\nabla_{||0}j_{||0} + \nabla\times\left(\frac{\mathbf{b}}{B}\right)\cdot\nabla P = 0
\label{eq:diveqj}
\end{equation}
From this the parallel current can be calculated, and compared with the parallel current
calculated from the input $f=R\Bt$ and pressure $P$ profiles:
\begin{equation}
\mu_0j_{||} = -B\deriv{f}{\psi} - \mu_0 \frac{f}{B}\deriv{P}{\psi} 
\label{eq:inputcurrent}
\end{equation}
For the MAST equilibrium shown in figure~\ref{fig:mast_mesh} and figure~\ref{fig:forcebalance} the parallel current
calculated from the curvature and pressure gradient (equation~\ref{eq:diveqj}) is shown in figure~\ref{fig:jparcomparison}, and compared to the current given by equation~\ref{eq:inputcurrent}.

These comparisons between quantities such as $J_{||}$ and 
curvature calculated in multiple ways allow an assessment of the quality of the input equilibrium. If the values obtained are inconsistent, then a higher accuracy input can be generated using a free boundary equilibrium solver such as EFIT or CORSICA~\cite{tarditi1996}. The result is a robust process for generating BOUT++ input grids from tokamak experimental equilibria, which has been successfully used for a range of devices.

\section{Conclusions}
\label{sec:conclusions}

Recent improvements to the BOUT++ simulation framework have been summarised. Since its original release~\cite{Dudson2009}, 
BOUT++ has been adopted by a growing number of users, who have extended its capabilities in a number of ways: the structure
of the code has been improved; more complex elliptic and parabolic equations can be solved, through coupling to the PETSc
library and built-in implementations; the input and output tools have been improved to enable experimental or theoretical
equilibria and profiles to be imported into BOUT++.

A preconditioning scheme has been presented which enables simulations to be performed without invoking the Boussinesq
approximation in the vorticity. This method can now be used to perform more accurate simulations of tokamak edge turbulence and ELMs, the exploration of which is the subject of future work.
More general elliptic and parabolic solvers for preconditioning and vorticity inversion problems in 3D are currently under
development, and will be presented elsewhere.

Soft scaling studies have been performed for a parabolic solver along equilibrium field-lines. We find that although 
the scaling with problem size is good (being based on the O(n) serial Thomas algorithm), the scaling with processor number
is quite poor due to the global gather and scatter operations used. Improving this will be a subject of future work.

\section*{Acknowledgements}

This work was funded by EPSRC grant EP/K006940/1 using HECToR computing
resources through the Plasma HEC consortium grant EP/L000237/1. 
EURATOM Mobility support is gratefully acknowledged. The views and opinions
expressed herein do not necessarily reflect those of the European Commission.

\bibliography{report}

\begin{thebibliography}{10}

\bibitem{goswami2001}
Goswami, R. et~al.,
\newblock Physics of Plasmas {\bf 8} (2001) 857.

\bibitem{fundamenski2001}
Fundamenski, W. et~al.,
\newblock J. Nucl. Materials {\bf 290-293} (2001) 593.

\bibitem{nakamura2011}
Nakamura, M.,
\newblock J. Nucl. Materials {\bf 415} (2011) S553.

\bibitem{rognlien1992}
Rognlien, T.~D. et~al.,
\newblock J. Nucl. Materials {\bf 196–198} (1992) 347.

\bibitem{schneider2006}
Schneider, R. et~al.,
\newblock Contrib. Plasma Phys. {\bf 46} (2006) 3.

\bibitem{kawashima2006}
Kawashima et~al.,
\newblock Plasma Fusion Res. {\bf 1} (2006) 31.

\bibitem{dippolito2011}
D'Ippolito, D.~A., Myra, J.~R., and Zweben, S.~J.,
\newblock Physics of Plasmas {\bf 18} (2011) 060501.

\bibitem{garcia-2007}
Garcia, O.~E. et~al.,
\newblock J. Nucl. Materials {\bf 363} (2007) 575.

\bibitem{naulin-2007}
Naulin, V. et~al.,
\newblock J. Nucl. Materials {\bf 24} (2007) 363.

\bibitem{ghendrih2012}
Gendrih, P. et~al.,
\newblock {J. Phys.: Conf. Ser.} {\bf 401} (2012) 012007.

\bibitem{ricci2012}
Ricci, P. et~al.,
\newblock Plasma Phys. Control. Fusion {\bf 54} (2012).

\bibitem{ricci2013}
Ricci, P. and Rogers, B.~N.,
\newblock Physics of Plasmas {\bf 20} (2013) 010702.

\bibitem{tamain2010}
Tamain, P. et~al.,
\newblock J. Comput. Phys. {\bf 229} (2010) 361.

\bibitem{Dudson2009}
Dudson, B.~D. et~al.,
\newblock Comp. Phys. Comm. {\bf 180} (2009) 1467.

\bibitem{xu-1999}
Xu, X.~Q. et~al.,
\newblock J. Nucl. Materials {\bf 266-269} (1999) 993.

\bibitem{xu-2000}
Xu, X.~Q. et~al.,
\newblock Nucl. Fusion {\bf 40} (2000) 731.

\bibitem{umansky-2006}
Umansky, M.~V., Rognlien, T.~D., Xu, X.~Q., Dudson, B.~D., and Kirk, A.,
\newblock {Modelling of edge plasma turbulence in a spherical tokamak},
\newblock 2006.

\bibitem{xu-2008}
Xu, X.~Q., Umansky, M.~V., Dudson, B., and Snyder, P.~B.,
\newblock Comm. in Comput. Phys. {\bf 4} (2008) pp. 949.

\bibitem{xu2010}
Xu, X.~Q. et~al.,
\newblock Phys. Rev. Lett. {\bf 105} (2010) 175005.

\bibitem{xia2012}
Xia, T.~Y., Xu, X.~Q., Dudson, B.~D., and Li, J.,
\newblock Contrib. Plasma Phys. {\bf 52} (2012) 353.

\bibitem{xi2013}
Xi, P.~W., Xu, X.~Q., Xia, T.~Y., Nevins, W.~M., and Kim, S.~S.,
\newblock Nucl. Fusion {\bf 53} (2013) 113020.

\bibitem{friedman2012}
Friedman, B., Carter, T.~A., Umansky, M.~V., Schaffner, D., and Dudson, D.,
\newblock Physics of Plasmas {\bf 19} (2012) 102307.

\bibitem{angus2012prl}
Angus, J.~R., Umansky, M.~V., and Krasheninnikov, S.~I.,
\newblock Phys. Rev. Lett. {\bf 108} (2012) 215002.

\bibitem{angus2012cpp}
Angus, J.~R., Umansky, M.~V., and Krasheninnikov, S.~I.,
\newblock Contrib. Plasma Phys. {\bf 52} (2012) 348.

\bibitem{walkden2013}
Walkden, N.~R., Dudson, B.~D., and Fishpool, G.,
\newblock Plasma Phys. Control. Fusion {\bf 55} (2013) 105005.

\bibitem{efficient}
Balay, S., Gropp, W.~D., McInnes, L.~C., and Smith, B.~F.,
\newblock in {\em Modern Software Tools in Scientific Computing}, edited by
  Arge, E., Bruaset, A.~M., and Langtangen, H.~P., pages 163--202, Birkhauser
  Press, 1997.

\bibitem{roache1998}
Roache, P.~J.,
\newblock {\em Verification and Validation in Computational Science and
  Engineering},
\newblock Hermosa Publishers, Albuquerque NM, 1998.

\bibitem{salari2000}
Salari, K. and Knupp, P.,
\newblock Code verification by the method of manufactured solutions,
\newblock Technical Report SAND2000-1444, Sandia National Laboratories, 2000.

\bibitem{gamma1995}
Gamma, E., Helm, R., Johnson, R., and Vlissides, J.,
\newblock {\em Design Patterns: Elements of Reusable Object-Oriented Software},
\newblock Addison-Wesley, 1995.

\bibitem{knepley2012}
Knepley, M.~G.,
\newblock arXiv:1209.1711  (2012).

\bibitem{xu2013}
Xu, X. et~al.,
\newblock Physics of Plasmas {\bf 20} (2013) 056113.

\bibitem{dimits2013}
Dimits, A.~M., Joseph, I., and Umansky, M.~V.,
\newblock Physics of Plasmas {\bf 21} (2013) 055907.

\bibitem{hazeltine-2003}
Hazeltine, R.~D. and Meiss, J.~D.,
\newblock {\em {Plasma Confinement}},
\newblock Dover publications, 2003.

\bibitem{catto-2004}
Catto, P.~J. and Simakov, A.~N.,
\newblock Physics of Plasmas {\bf 11} (2004) pp. 90.

\bibitem{ottaviani1999}
Ottaviani, M. and Manfredi, G.,
\newblock Physics of Plasmas {\bf 6} (1999) 3267.

\bibitem{beer1997}
Beer, M.~A. et~al.,
\newblock Physics of Plasmas {\bf 4} (1997) 1792.

\bibitem{scott-2005}
Scott, B.,
\newblock Physics of Plasmas {\bf 12} (2005) 102307.

\bibitem{iserles2009}
Iserles, A.,
\newblock {\em A First Course in the Numerical Analysis of Differential
  Equations},
\newblock Cambridge University Press, 2009,
\newblock ISBN: 978-0-521-73490-5.

\bibitem{yu2006}
Yu, G.~Q., Krasheninnikov, S.~I., and Guzdar, P.~N.,
\newblock Physics of Plasmas {\bf 13} (2006) 042508.

\bibitem{scott-2002}
Scott, B.~D.,
\newblock New J. Physics {\bf 4} (2002) 52.1.

\bibitem{scott03}
Scott, B.,
\newblock Plasma Phys. Control. Fusion {\bf 45} (2003) A385.

\bibitem{scott2005-arxiv}
Scott, B.~D.,
\newblock arXiv:physics/0501124  (2005).

\bibitem{angus2014}
Angus, J.~R. and Umansky, M.~V.,
\newblock Physics of Plasmas {\bf 21} (2014) 012514.

\bibitem{petsc-user-ref}
Balay, S. et~al.,
\newblock Technical Report ANL-95/11 - Revision 3.1, Argonne National
  Laboratory, 2010.

\bibitem{austin2004}
Austin, T.~M., Berndt, M., and Moulton, J.~D.,
\newblock Tech.Rep. LA-UR 03-4149, LANL, USA, 2004.

\bibitem{dudson2012}
Dudson, B., Farley, S., and Curfmann~McInnes, L.,
\newblock arXiv:1209.2054  (2012).

\bibitem{hammett90}
Hammett, G.~W. and Perkins, F.~W.,
\newblock Phys. Rev. Lett. {\bf 64} (1990) 3019.

\bibitem{karniadakis1991}
Karniadakis, G.~E., Israeli, M., and Orszag, S.~A.,
\newblock J. Comput. Phys. {\bf 97} (1991) 414.

\bibitem{hindmarsh2005}
Hindmarsh, A.~C. et~al.,
\newblock {ACM} Transactions on Mathematical Software {\bf 31} (2005) 363.

\bibitem{ChaconKnollEtAl02}
Chac\'{o}n, L., Knoll, D.~A., and Finn, J.~M.,
\newblock J. Comput. Phys. {\bf 178} (2002) 15.

\bibitem{wesson-1997}
Wesson, J.~A., editor,
\newblock {\em Tokamaks},
\newblock Clarendon Press, 2 edition, 1997.

\bibitem{CTPP:CTPP201210038}
Tskhakaya, D.,
\newblock Contrib. Plasma Phys. {\bf 52} (2012) 490.

\bibitem{CTPP:CTPP200810015}
Tskhakaya, D. et~al.,
\newblock Contrib. Plasma Phys. {\bf 48} (2008) 89.

\bibitem{ji:022312}
Ji, J.~Y., Held, E.~D., and Sovinec, C.~R.,
\newblock Physics of Plasmas {\bf 16} (2009) 022312.

\bibitem{omotani2013}
Omotani, J.~T. and Dudson, B.~D.,
\newblock Plasma Phys. Control. Fusion {\bf 55} (2013) 055009.

\bibitem{compilers}
Aho, A.~V., Lam, M.~S., Sethi, R., and Ullman, J.~D.,
\newblock {\em Compilers: Principles, Techniques, and Tools},
\newblock Addison-Wesley, 2006.

\bibitem{llvmtutorial}
{LLVM Project},
\newblock {Implementing a language with LLVM},
\newblock http://llvm.org/docs/tutorial/.

\bibitem{scipy}
Jones, E. et~al.,
\newblock {SciPy}: Open source scientific tools for {Python}, 2001--.

\bibitem{rognlien-2002}
Rognlien, T.~D., Xu, X.~Q., and Hindmarsh, A.~C.,
\newblock J. Comput. Phys. {\bf 175} (2002) 249.

\bibitem{marchand1996}
Marchand, R. and Dumberry, M.,
\newblock Comp. Phys. Comm. {\bf 96} (1996) 232.

\bibitem{ma2014}
Ma, J.~F., Xu, X.~Q., and Dudson, B.~D.,
\newblock Nucl. Fusion {\bf 54} (2014) 033011.

\bibitem{tarditi1996}
Tarditi, A. et~al.,
\newblock Contrib. Plasma Phys. {\bf 36} (1996) 132.

\end{thebibliography}
\bibliographystyle{cpc}

\end{document}